\documentclass[conference]{IEEEtran}
\IEEEoverridecommandlockouts
\usepackage{cite}
\usepackage{amsmath,amssymb,amsfonts}
\usepackage{algorithmic}
\usepackage{graphicx}
\usepackage{textcomp}
\usepackage{xcolor}
\usepackage[inline]{enumitem}
\def\BibTeX{{\rm B\kern-.05em{\sc i\kern-.025em b}\kern-.08em
    T\kern-.1667em\lower.7ex\hbox{E}\kern-.125emX}}
\usepackage{enumitem}
\usepackage{xcolor,comment}

\begin{document}

\title{Who Needs MLOps: What Data Scientists Seek to Accomplish and How Can MLOps Help?}

\author{
\IEEEauthorblockN{Sasu M{\"a}kinen}
\IEEEauthorblockA{
\textit{University of Helsinki}\\
Helsinki, Finland \\
sasu.makinen@helsinki.fi}
\and
\IEEEauthorblockN{Henrik Skogstr{\"o}m}
\IEEEauthorblockA{
\textit{Valohai}\\
Turku, Finland \\
henrik@valohai.com}\\
\and
\IEEEauthorblockN{Eero Laaksonen}
\IEEEauthorblockA{
\textit{Valohai}\\
Turku, Finland \\
eero@valohai.com}\\
\and
\IEEEauthorblockN{Tommi Mikkonen}
\IEEEauthorblockA{
\textit{University of Helsinki}\\
Helsinki, Finland \\
tommi.mikkonen@helsinki.fi}
}

\maketitle

\begin{abstract}
Following continuous software engineering practices, there has been an increasing interest in rapid deployment of machine learning (ML) features, called MLOps. In this paper, we study the importance of MLOps in the context of data scientists' daily activities, based on a survey where we collected responses from 331 professionals from 63 different countries in ML domain, indicating on what they were working on in the last three months. Based on the results, up to 40\% respondents say that they work with both models and infrastructure; the majority of the work revolves around relational and time series data; and the largest categories of problems to be solved are predictive analysis, time series data, and computer vision. The biggest perceived problems revolve around data, although there is some awareness of problems related to deploying models to production and related procedures. To hypothesise, we believe that organisations represented in the survey can be divided to three categories -- (i) figuring out how to best use data; (ii) focusing on building the first models and getting them to production; and (iii) managing several models, their versions and training datasets, as well as retraining and frequent deployment of retrained models. In the results, the majority of respondents are in category (i) or (ii), focusing on data and models; however the benefits of MLOps only emerge in category (iii) when there is a need for frequent retraining and redeployment.  Hence, setting up an MLOps pipeline is a natural step to take, when an organization takes the step from ML as a proof-of-concept to ML as a part of nominal activities.
\end{abstract}


\section{Introduction}

Continuous software engineering practices \cite{fitzgerald2017continuous}, such as DevOps \cite{debois2011devops,bass2015devops}, have been rapidly adopted by software development organizations in their operations. This has resulted in an ability to deploy new features to use whenever their development is completed, which in turn means that systems are updated frequently. To achieve this goal, there is a growing need for collaboration across the organization -- in the case of DevOps, this mainly involves developers and operators \cite{davis2016effective}, but for more complex situations, involving security or regulations, for instance, also other participants may be needed.

More recently, there has been an increasing interest in rapid deployment of machine learning (ML) features. The practice of continuous delivery of ML is called MLOps. it mimics DevOps practices but introduces additional actions that are specific to ML. This seems only natural, as in many cases ML features are embedded into a larger software system that hosts, provides access to, and monitors ML features. In fact, often the model, which can be the core of the application, is just a small part of the whole software system, so the interplay between the model and the rest of the software and context is essential~\cite{sculley2015hidden}. 

In this paper, we study the importance of MLOps in the context of data scientists' daily activities. As concrete results, we present a survey where we collected responses from 331 professionals working in the ML domain, indicating on what they were working on in the last three months, and study the role MLOps might play in their daily activities.

The rest of this paper is structured as follows. In Section 2, we present the background and motivation of this work. In Section 3, we introduce the survey, its questions and goals, and the results of the survey. In Section 4, we discuss the implications of the survey findings. Towards the end of the paper, in Section 5, we draw some final conclusions.

\section{Background and Motivation}

Building on the success of continuous software development approaches \cite{fitzgerald2014continuous,fitzgerald2017continuous}, in particular DevOps \cite{lwakatare2015dimensions}, it has become desirable to deploy machine learning (ML) components in real time, too. To this end, MLOps refers advocating automation and monitoring at all steps of ML system development and deployment, including integration, testing, releasing, deployment and infrastructure management.


To understand the challenges related to MLOps, let us first explain the steps necessary to train and deploy ML modules \cite{SWQD21}. As the starting point, data must be available for training. There are various somewhat established ways of dividing the data to training, testing, and cross-validation sets. Then, an ML model has to be selected, together with its hyperparameters. 
Next, the model is trained with the training data. During the training phase, the system is iteratively adjusted so that the output has a good match with the ``right answers'' in the training material. This trained model can also be validated with different data. If this validation is successful -- with any criteria we decide to use -- the model is ready for deployment, similarly to any other component. 

Once deployed, ML related features need monitoring, like any other deployed feature. However, monitoring in the context of ML must take into account inherent ML related features, such as biases and drift that may emerge over time. In addition, there are techniques that allow improving the model on the fly, while it is being used. Therefore, the monitoring system must take these special needs into account.

To summarize, phases in ML that precede completing the ML model seem to be waterfallish in their nature, whereas operationalizing the model to a larger whole follows the practices associated with conventional software. This observation is in line with our previous study, where data scientists say that to a degree, early phases of a data science project are about understanding the data, and only after that, when deploying the results to production, it is desirable to incorporate the outcome into a bigger whole \cite{aho2020demystifying}.  

\begin{figure}[htb]
    \centering
    \includegraphics[width=0.99\columnwidth]{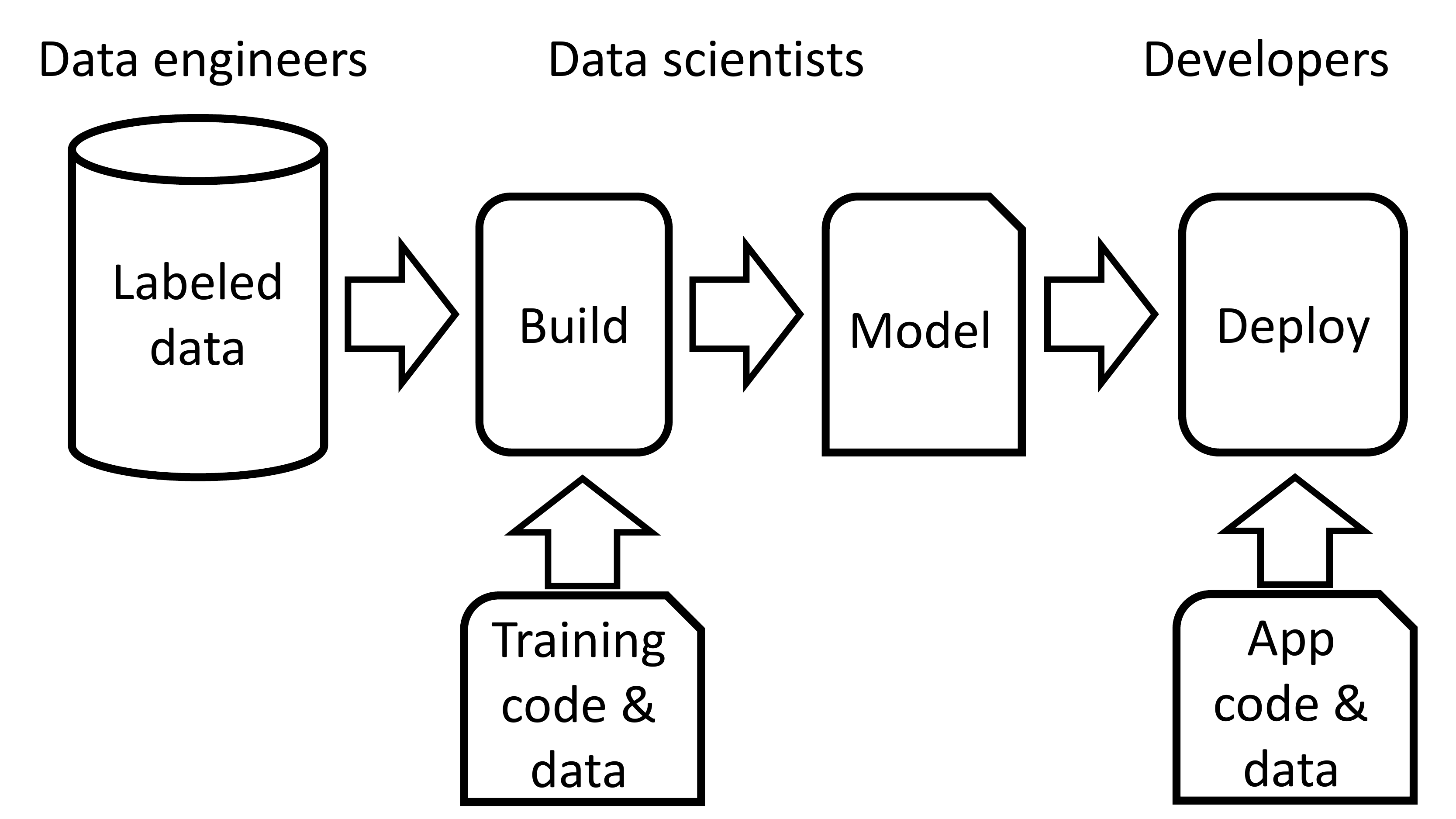}
    \caption{A simplified CD4ML pipeline and its artifacts, with conventional task allocation for data engineers, data scientists, and application developers \cite{cd4ml}.}
    \label{fig:cd4ml}
\end{figure}

So far, while numerous designs exist, one of the most established formalization of MLOps is Continuous Delivery for Machine Learning (CD4ML) \cite{cd4ml}. CD4ML is an approach proposed by ThoughtWorks, and it automates the lifecycle of machine learning applications from end-to-end. In the proposed approach, a cross-functional team produces machine learning applications based on code, data, and models in small and safe increments that can be reproduced and reliably released at any time, in short adaptation cycles. The approach contains three distinct steps (Figure \ref{fig:cd4ml}): 
\begin{enumerate*}
\item identify and prepare the data for training, 
\item experimenting with different models to find the best performing candidate, and \item deploying and using the selected model in production. 
\end{enumerate*}
Of these, data operations are executed by data engineers; model building and experimentation is run by data scientists, and model deployment and use is done by application developers.

\section{Survey}

To understand the difficulties data scientists are having in their daily work, an online survey was created. The \textit{State of ML 2020 survey} consisted of a number of questions, addressing the background and the  activities performed by the data scientists.  The surveyed demographics were country of origin, title, team role, number of employees, and industry. As for the actual content, we surveyed  ML related problems that the respondent is trying to solve,  the kind of data they are working with, and what they are trying to accomplish in the next three months.  In addition, the survey had a section for rating the relevance of following obstacles: 
\begin{itemize}[label=--]
    \item Lack of data	
    \item Messiness of data	
    \item Accessibility of data	
    \item Not enough data scientists	
    \item Not enough engineers/DevOps personnel	
    \item Not enough budget for purchases (solutions, computing)	
    \item Difficulty of developing an effective model	
    \item Difficulty of using cloud resources for training	
    \item Difficulty of building training pipelines	
    \item Difficulty of deploying models	
    \item Difficulty of tracking and comparing experiments	
    \item Difficulty of collaborating on projects	
    \item Lack of version control	
    \item Lack of executive buy-in	
    \item Difficult regulatory environment	
    \item Unclear or unrealistic expectations.
\end{itemize}
Some of the questions were open-ended, whereas some had a fixed set of answers to select from.

\subsection{Data Collection}

The \textit{State of ML 2020 survey} was conducted online in May 2020. We received 331 responses to the survey from 63 different countries. The survey respondents were sourced through three different channels: the Valohai community, the Data Science Weekly newsletter, and LinkedIn. The Valohai community consists of the Valohai platform users and a broad section of the data science community as Valohai publishes expert articles on various machine learning topics. The Data Science Weekly newsletter is an email newsletter with 43000 subscribers globally interested in data science. In LinkedIn, both advertisements and outreach were utilized to reach survey participants. The criteria was that the respondents would have data science or machine learning related title, such as data scientist, machine learning engineer, data engineer, or head of data science.

\subsection{Results}

\begin{figure}[t]
    \centering
    \includegraphics[width=\columnwidth]{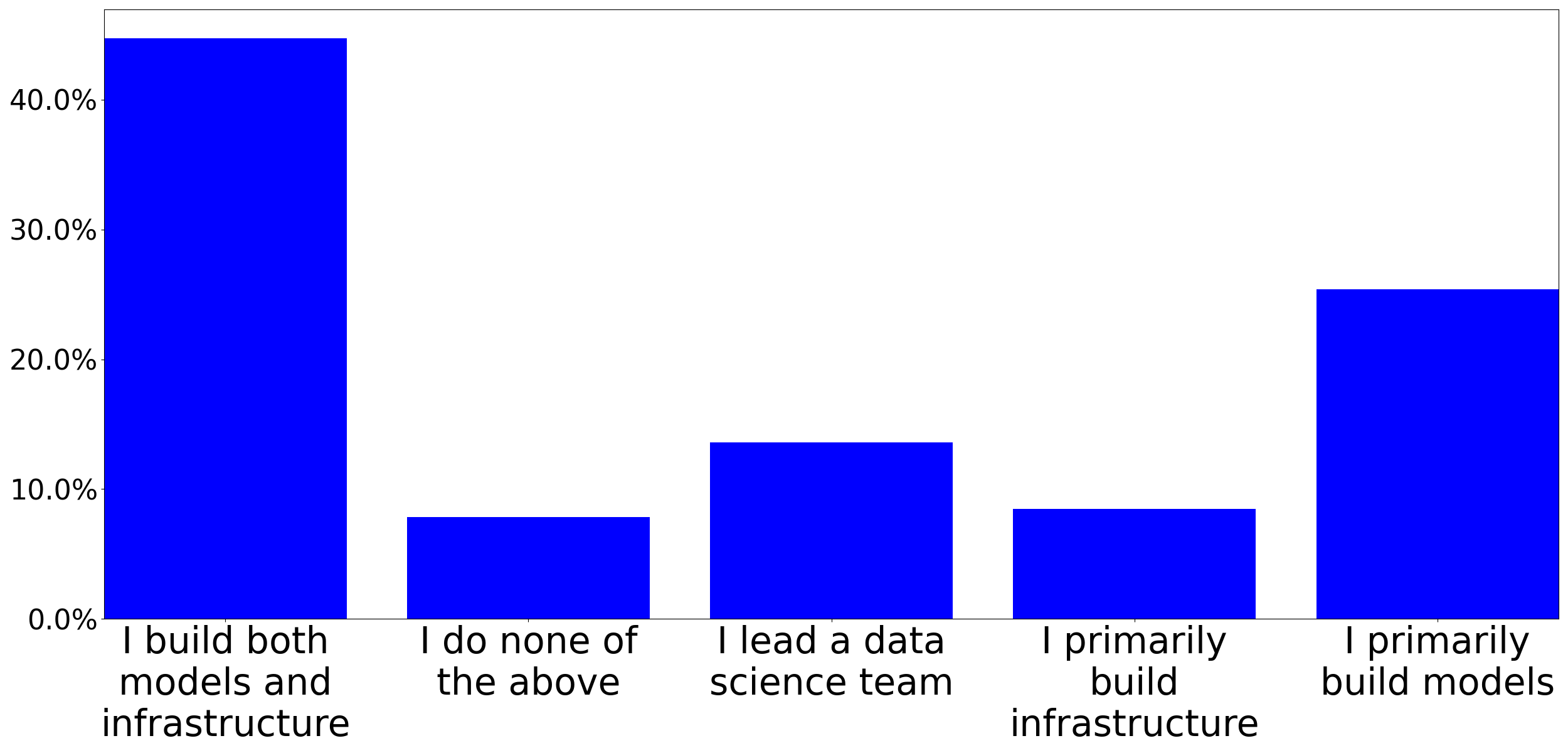}
    \caption{What is your role in your team?}
    \label{fig:roles}
\end{figure}

\begin{figure}[t]
    \centering
    \includegraphics[width=\columnwidth]{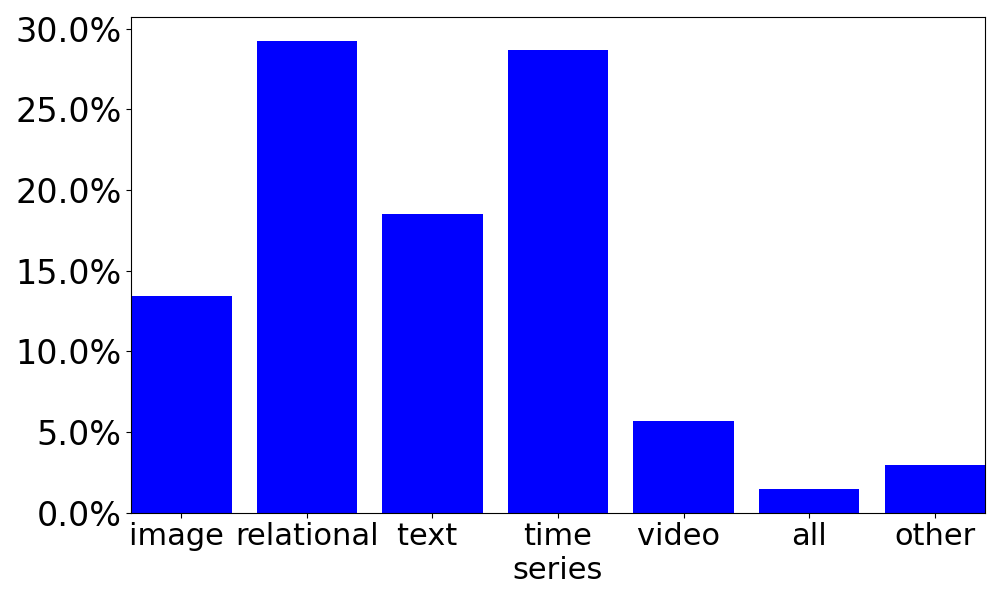}
    \caption{What kind of data are you working with?}
    \label{fig:kind_of_data}
\end{figure}

\begin{figure}[t]
    \centering
    \includegraphics[width=\columnwidth]{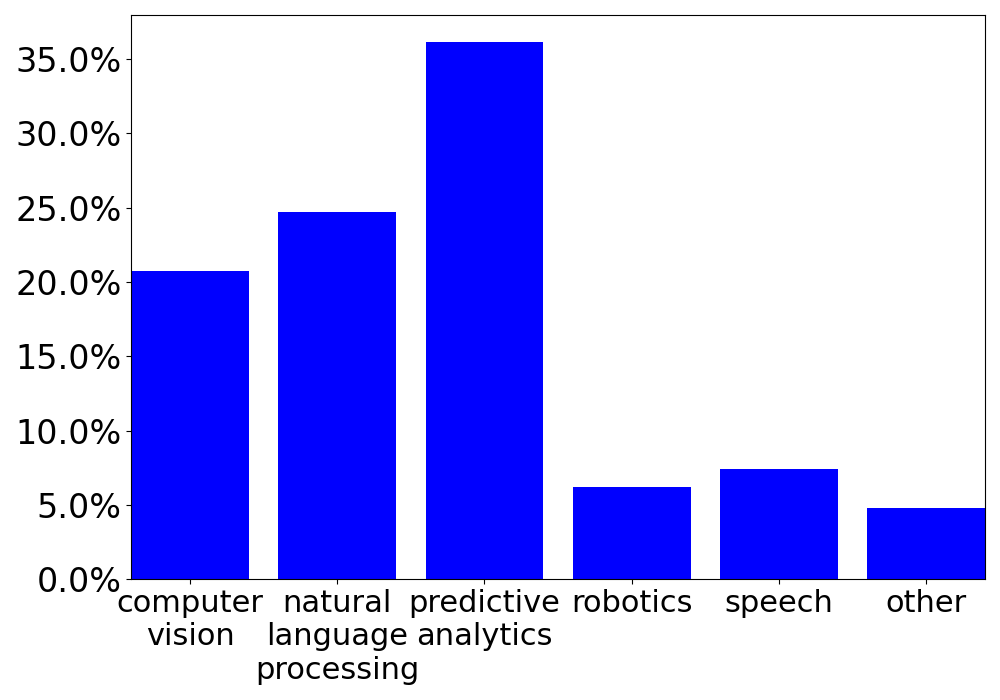}
    \caption{What kind of problems is your company solving with machine learning?}
    \label{fig:problems_solving}
\end{figure}

Responses were received from all other continents but Antarctic, with the main bias being that Finland in particular and Europe in general are slightly over-represented in the responses. Respondents represent a mix of different types of organizations, ranging from startups to large enterprises, and also the teams where the respondents work are of varying size. We have distilled key information from the survey to Figures \ref{fig:roles}, \ref{fig:kind_of_data}, and \ref{fig:problems_solving}, addressing the role of the respondent in the team, data involved, and the problems that the respondent tries to solve with ML. Perhaps surprisingly, up to 40\% respondents say that they work with both models and infrastructure; less surprisingly, the majority of the work revolves around relational and time series data.  As for the problems to be solved, the largest categories are predictive analysis, time series data, and computer vision.

\begin{figure}[t]
    \centering
    \includegraphics[width=\columnwidth]{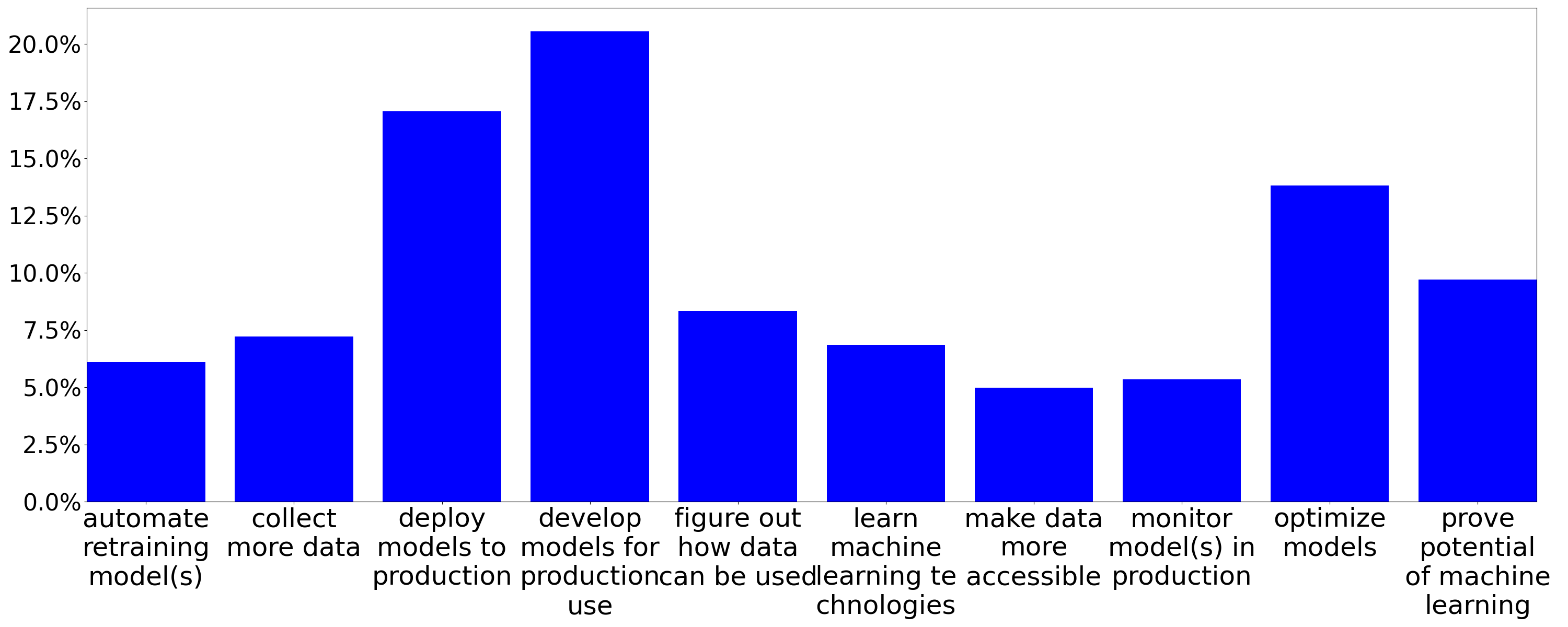}
    \caption{What are you trying to accomplish in the next 3 months?}
    \label{fig:next_3_months}
\end{figure}

In addition, to understand the goals of the work, we asked a more detailed question about short-term activities, to be completed within next 3 months. Answers to this question are summarized in Figure \ref{fig:next_3_months}. In general, the answers show wide divergence in the actions, with four top goals being (in descending order) developing models for production use, deploying models to production, optimizing models, and proving the potential of machine learning.

\begin{figure}[t]
     \centering
    \includegraphics[width=.99\columnwidth]{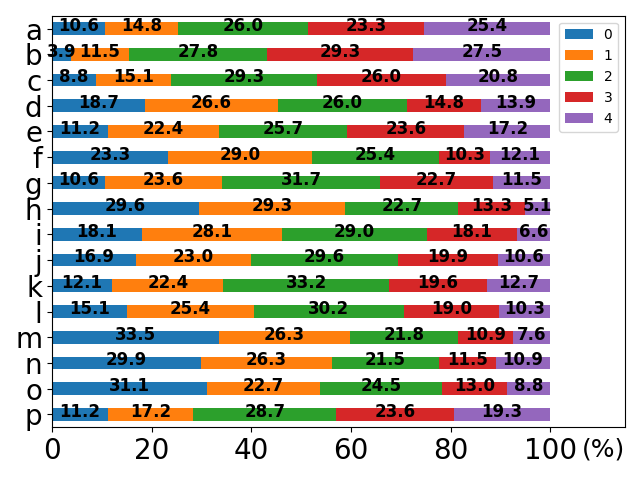}

    \caption{Perceived challenges based on survey data. Legend:
    (a) Lack of data;
    (b) Messiness of data;	
    (c) Accessibility of data;	
    (d) Not enough data scientists;	
    (e) Not enough engineers/DevOps personnel;	
    (f) Not enough budget for purchases (solutions, computing);	
    (g) Difficulty of developing an effective model;	
    (h) Difficulty of using cloud resources for training;	
    (i) Difficulty of building training pipelines;	
    (j) Difficulty of deploying models;	
    (k) Difficulty of tracking and comparing experiments;	
    (l) Difficulty of collaborating on projects;	
    (m) Lack of version control;	
    (n) Lack of executive buy-in;	
    (o) Difficult regulatory environment;	
    (p) Unclear or unrealistic expectations.
    }
    \label{fig:challenges}
\end{figure}

Finally, the key challenges that the companies identified in the context in ML are visualized in Figure \ref{fig:challenges}. Here, the majority of respondents identify data related issues as the biggest problems, with messiness of data being the prime one, followed by the lack of data and accessibility of data. However, unclear or unrealistic expectations are also considered problematic.

\subsection{Analysis}

To analyse who needs MLOps based on the answers, we classified organizations that do machine learning in terms of their ML maturity. These resulted in the following categorization:
\begin{itemize}
\item Data-centric: the organization is figuring out how to manage and utilize data.
\item Model-centric: the organization is figuring out how to build their first model and get that to production.
\item Pipeline-centric: the organization has models in production, and they are increasingly business-critical. The organization is constantly considering how to scale and do continuous development while at the same time maintaining the quality of production models.
\end{itemize}
Our hypothesis is that in general, organizations start with data-centric setup and then advance to model-centric mode, when they have solved issues that are related to data engineering. Then, when an organization masters building models, it can turn to making ML a standard operational procedure, which requires automation in the form of pipelines -- in essence, MLOps.

Based on the overall survey results, most respondents represent model-centric organizations in terms of their operations, with the main focus placed on developing a model and pushing it to production. In addition, there is a considerable number of respondents who struggle with data engineering issues, with pipeline-centric considerations being the minority. 

When placing the focus only on respondents representing organizations that are automating the retraining of models -- pipeline-centric ones, with 22 respondents indicating this as their main concern -- data-centric tasks, including, for instance, how to enable data utilization and prove its value, are no longer a central issue. Such organizations, pioneers in the ML industry, have some shared characteristics. These are listed in the following:
\begin{itemize}
\item A tendency to have a data science team (11-30 persons), indicating a strategic commitment at organizational level.
\item A tendency to have individuals who are concerned about both infrastructure and models, in contrast to just focusing on models or data as such and overlooking the delivery and deployment part.
\item  Scale-ups (151-500 employees) and Enterprises (1001+) are the most pipeline-centric organizations.
\end{itemize}
We trust that these organizations are also the ones where MLOps can bring the most value at the level of daily operations, removing the burden of building training pipelines.

\section{Discussion}

In general, the survey shows that ML is moving away from one-man proof-of-concepts, described in \cite{aho2020demystifying}, and advancing towards more mature setups where a team of developers work together in ML development. Furthermore, there is a growing interest in infrastructure issues, although frequent deployment is not yet a key concern in most organizations.

Looking at respondents roles, it seems that the skills of a data scientist are expanding from data science and models to other domains, in particular ML infrastructure and deployment. At present it is unclear how narrow or wide should one’s area of expertise be -- in analogy to web development \cite{northwood2018full}, the "full stack" of ML remains largely undefined. 

\section{Conclusions}

Similarly to DevOps for traditional software, the ability to continuously deliver ML software -- so-called MLOps -- is becoming a requirement for companies that apply ML in production. However, DevOps tools cannot be simply apply plug-and-play approach for MLOps, because there are several characteristics that are different. In particular, data and its complexities and model construction and training are new concerns that introduce new types of computational requirements. Furthermore, the complexity of maintaining coherence and quality increases as the number of models grows, requiring a systematic approach to versioning also with respect to models and data sets.

In this paper, we have studied the state of ML to understand to what extent MLOps is needed in today's ML operations today, and to what extent data scientists are still struggling with data and modeling issues only, without considering their deployment. The responses show that while the majority of respondents still work with data and models, the pioneers are advancing to MLOps arena. In general, these pioneers also show certain maturity in their ML operations, with larger teams and more infrastructure-oriented mindset than in other organizations.

\section*{Acknowledgement}

The authors would sincerely like to thank Valohai (https://valohai.com/) for an access to survey data, and Business Finland (project AIGA -- AI Governance and Auditing) for supporting this research.

\bibliographystyle{plain}
\bibliography{bib}

\end{document}